\documentstyle[aps,epsfig,eqsecnum]{revtex}
\newcommand{\bol}[1]{\mbox{\boldmath $#1$}}
\newcommand{\be}{\begin{eqnarray}}
\newcommand{\ee}{\end{eqnarray}}

\newcommand{\ket}[1]{\mbox{$\mid #1\,\rangle$}}

\newcommand{\real}{\mbox{{\rm I\hspace{-2truemm} R}}}

\newcommand{\id}{\mbox{{\rm 1\hspace{-2truemm} I}}}
\begin{document}
\draft
\title{Three dimensional gravity from ISO(2,1) coset models}
\author{R. Casadio}
\address{Dipartimento di Fisica, Universit\`a di Bologna, and \\
I.N.F.N., Sezione di Bologna, \\
via Irnerio 46, 40126 Bologna, Italy}
\author{B. Harms}
\address{Department of Physics and Astronomy,
The University of Alabama\\
Box 870324, Tuscaloosa, AL 35487-0324}
\date{\today}
\maketitle
\begin{abstract}
Starting from a WZWN action in the $ISO(2,1)$ Poincar\'e group which
describes a bosonized spinning string in $2+1$ Minkowski space-time,
we show that a sequence of non-trivial compactifications leads to
the description of a spinless string which moves in a (linear dilaton)
vacuum, AdS$_3$ or BTZ black hole background.
Other solutions are also obtained and their T-duals analyzed.
\end{abstract}
%
%
\section{Introduction}
\par
The observational evidence for the existence of black holes in nature
is now very strong.
The data support the existence of both supermassive black holes at
the centers of galaxies and smaller (a few solar masses up
to a few tens of solar masses) black holes in binary systems
\cite{bh}.
The best candidate for a unified theory of all the physical phenomena
observed so far, including black holes, is string theory \cite{gsw}
and, indeed, several black $p$-brane solutions have been found
in various space-time dimensions in the low energy limit of this theory
(for a review see, {\em e.g.}, Ref.~\cite{duff}).
However, only one black hole \cite{horowitz} is known to exist in the
three-dimensional low energy limit of string theory and it coincides with
the only known black hole in three-dimensional Einstein
gravity \cite{witten2}, to wit the (BTZ) black hole of Ba\~nados,
Teitelboim and Zanelli \cite{BTZ} (see also Ref.~\cite{brill}).
\par
Although the BTZ black hole is not useful as a global description of
real black holes (for example, the curvature of the BTZ black hole is
constant and there are no gravitational waves in three dimensions),
it does provide a manageable model of string propagation on a black
background in which an infinite number of propagating modes is present.
The Green's function for this black hole can be constructed, and the
quantum stress tensor can be calculated from it \cite{steif}.
This system has also been used to study such problems as the quantization
of a string on a black hole background (see \cite{satoh} and Refs.
therein).
\par
Recently, the theoretical interest in the BTZ black hole has also been
raised by the conjectured AdS/CFT correspondence \cite{maldacena},
according to which all the relevant quantities of the gravitational
field theory in the bulk of the anti-de~Sitter (AdS) space-time (or any
space-time with a time-like boundary) can be described in terms of
a conformal field theory (CFT) on the boundary.
Thus, by applying this conjecture to the black $p$-branes there is some
hope of describing the complete evolution of a black hole, from its
formation \cite{sen} to the evaporation \cite{hawking}, and solve the
riddle of its final fate (see Ref.~\cite{birrell} for a list of still
unanswered questions).
However, it is not clear whether the AdS/CFT correspondence extends
beyond perturbation theory on a given background manifold \cite{proof}
as the solution of the black hole problem would require in order
to compute the backreaction of the evaporation radiation on the
geometry \cite{mfd}.
\par
Because of the usefulness of three-dimensional black holes as
prototypes for four-dimensional black holes in string theory,
a search for a second exact, three-dimensional black hole in string
theory would seem to be a worthwhile pursuit, especially if one could
be found which has non-negative curvature.
This work describes our attempt to obtain such a solution, starting
from Wess-Zumino-Witten-Novikov (WZWN) model in the Poincar\'e
group $ISO(2,1)$ \cite{stern}.
Our procedure for obtaining a three-dimensional metric is to promote
the six parameters of the $ISO(2,1)$ group to space-time variables and
then to reduce space-time to three dimensions by various
compactifications.
After each compactification we investigate the symmetries of the
resulting model.
\par
A partial result has been obtained, since we can now show that the
string theory we start from can be compactified in such a way as to yield
either a (linear dilaton) vacuum and AdS$_3$ (or the BTZ black hole).
We also obtaine other solutions which contain a non-trivial dilaton field
and, thus, might be of interest for studying the evaporation.
\par
In Section~\ref{wzwn} we review the WZWN Poincar\'e action in three
dimensions and, in Section~\ref{coset}, its coset descendants
$ISO(2,1)/\real^n$ \cite{ch,ch2}.
In Section~\ref{time} we specialize to the case when $\real$ is the
translation in the time direction and, in Section~\ref{final}, we
further compactify to three-dimensional space-time in which we
recover the AdS (and BTZ) manifold.
In Section~\ref{other} we describe other solutions and their T-duals,
and finally comment on our results in Section~\ref{end}.
For the metric and other geometrical quantities we follow the convention
of Ref.~\cite{mtw}.
\section{ISO(2,1) WZWN Models}
\label{wzwn}
The WZWN construction starts with the $\sigma$-model action at level
$\kappa$ \cite{witten}
\be
S_{\sigma}={\kappa\over{4}}\int_{\partial {\cal M}} d^2\sigma\,
{\rm Tr}\,\left(g^{-1}\,\partial_+g\,g^{-1}\,\partial_-g\right)
-{\kappa\over{4}}\int_{\cal M} d^3\zeta\,
{\rm Tr}\,\left(g^{-1}\,\partial g\wedge g^{-1}\,\partial g
\wedge g^{-1}\,\partial g\right)
\ ,
\ee
where in the present case $g$ is an element of the Poincar\'e group
$ISO(2,1)$ and $\sigma^\pm=\tau\pm\sigma$ are light-cone coordinates
on the boundary $\partial {\cal M}$ of the three-dimensional manifold
${\cal M}$.
\par
The elements of $ISO(2,1)$ can be written using the notation
$g=\left(\Lambda,v\right)$,
where $\Lambda\in SO(2,1)$ and $v\in \real^3$.
Given the map $g:\  {\cal M}=D^2\times R\mapsto ISO(2,1)$ from the
two-dimensional disc$\times$time to $ISO(2,1)$,
the action can be written entirely on the boundary $\real\times S^1$
and it describes a closed bosonized spinning string moving in 2+1
Minkowski space-time with coordinates $v^i$ \cite{stern},
\be
S=-{\kappa\over 4}\,\int_{\partial {\cal M}} d^2\sigma\,
\epsilon^{ijk}\,
\left(\partial_+\Lambda\,\Lambda^{-1}\right)_{ij}\,
\partial_- v_k
\ .
\label{S}
\ee
where $\epsilon^{ijk}$ is the Levi-Civita symbol in three dimensions
and the metric tensor is $\eta_{ij}={\rm diag}\,[-1,+1,+1]$
($i,j,\ldots=0,1,2$).
\par
The basic property of the action $S$ is that it is invariant under
\be
g\mapsto g_{_L}(\sigma^+)\,g\,g_{_R}^{-1}(\sigma^-)
\ ,
\label{inv}
\ee
where $g_{_{L/R}}\in ISO(2,1)$, and also under the left and right
action of the group of diffeomorphisms of the world-sheet \cite{stern}.
Starting from this observation, the canonical structure of the model can
be computed by reverting to the ``chiral'' version of Eq.~(\ref{S}),
which is obtained by formally replacing $\sigma^+\to\tau\in\real$ and
$\sigma^-\to\sigma \in(0,2\,\pi)$ \cite{chiral}.
One then finds two sets of conserved current densities,
the first of which is given by
\be
&&P^i(\sigma)={\kappa\over 2}\,\epsilon^{ijk}\,\left(\Lambda^{-1}\,
\partial_\sigma\Lambda\right)_{jk}
\label{P-}\\
&&J^i(\sigma)=\kappa\,\left(\Lambda^{-1}\,\partial_\sigma v\right)^i
\ ,
\label{J-}
\ee
with Poisson brackets \cite{stern}
\be
&&\left\{P^i(\sigma),P^j(\sigma')\right\}=0
\label{P,P}
\\
&&\left\{J^i(\sigma),J^j(\sigma')\right\}=
-\epsilon_{ijk}\,J_k(\sigma)\,\delta(\sigma-\sigma')
\label{J,J}
\\
&&\left\{J^i(\sigma),P^j(\sigma')\right\}=
-\epsilon^{ijk}\,P_k(\sigma)\,\delta(\sigma-\sigma')
+\kappa\,\eta^{ij}\,{\partial\over\partial\sigma}\,\delta(\sigma-\sigma')
\ ,
\label{J,P}
\ee
and generate $L^*ISO(2,1)$, the Poincar\'e loop group with the
central extension given by the last term in Eq.~(\ref{J,P}).
This is the algebra of the right transformations in Eq.~(\ref{inv}),
since in the chiral picture $g_{_R}(\sigma^-)\to g_{_R}(\sigma)$ has
become a space-dependent transformation of the field $g$ on the
world-sheet.
The (time dependent) left chiral transformation,
$g_{_L}(\sigma^+)\to g_{_L}(\tau)$, in Eq.~(\ref{inv}) is now an
$ISO(2,1)$ invariance generated by the zero Fourier modes of the second
set of (weakly vanishing) current densities
\be
&&\bar P^i={\kappa\over 2}\,
\epsilon^{ijk}\,\left(\partial_\sigma\Lambda\,\Lambda^{-1}\right)_{jk}
\label{P+}
\\
&&\bar J^i=\kappa\,
\left[\Lambda\,\partial_\sigma(\Lambda^{-1}\,v)\right]^i
\ .
\label{J+}
\ee
The latter commute with $P^i$ and $J^i$ and have Poisson brackets
among themselves given by Eqs.~(\ref{P,P}), (\ref{J,J}) and
(\ref{J,P}) with a central extension opposite in sign \cite{stern}.
One then concludes that (half \cite{chiral}) the (classical) gauge
invariant phase space of the model is $L^*ISO(2,1)/ISO(2,1)$.
\par
As usual, one expects the Fourier modes of $P_i$ and $J_i$ (the
Kac-Moody generators) in turn yield a Virasoro algebra (for each chiral
sector) whose generators $L_n$ are obtained via the Sugawara
construction.
However, there is a potential problem in the quantum theory since the
standard highest weight construction \cite{gsw}, which would give a
central charge $c={\rm dim}\,ISO(2,1)=6$, fails to deliver unitary
representations because not all negative norm states are suppressed
by the conditions $\hat L_n\,\ket{phys}=0$ for all $n\ge 0$.
Spaces of positive norm states can instead be obtained by employing
the method of induced representations which yields the Virasoro
generators as Fourier modes of
\be
L(\sigma)={1\over\kappa}\,J^i(\sigma)\,P_i(\sigma)
\ ,
\ee
and a central charge $c=0$ for each chiral sector \cite{stern}.
In either case, the total central charge of the model, after adding
the ghost contribution \cite{bound,gsw}, is $c_T=c-26$ and one must
eventually add $26-c$ bosonic degrees of freedom in order to have a
quantum model which is free of anomaly.
\par
The action (\ref{S}) is also one of the two exceptional cases described
in Ref.~\cite{ch2}, where it was shown that, if one considers all
parameters of the six-dimensional Poincar\'e group as space-time
coordinates, then $S$ describes a spinless string moving on a curved
background with six-dimensional metric.
It was also proved that this action is in some sense unique, since no
generalization of the kind studied in Refs.~\cite{stern2,ch2} exists for
the Poincar\'e group in three dimensions.
\section{Coset Models}
\label{coset}
The action (\ref{S}) is not invariant under the local action
of any subgroup $H$ of $ISO(2,1)$ given by
\be
h\cdot g:\
g\mapsto h_{_L}(\sigma^-,\sigma^+)\,g\,h_{_R}^{-1}(\sigma^-,\sigma^+)
\ ,
\label{ac}
\ee
where now $h_{_{L/R}}=(\theta_{_{L/R}},y_{_{L/R}})\in H$,
due to the dependence of $h_{_L}$ on $\sigma^-$
and of $h_{_R}$ on $\sigma^+$.
However, $H$ can in general be promoted to a gauge symmetry of
the action by introducing suitable gauge fields
$A_\pm=\left(\omega_\pm,\xi_\pm\right)$ belonging to the Lie algebra
of $H$, and the corresponding covariant derivatives
$D_\pm=\partial_\pm+A_\pm$.
\par
In order that $ISO(2,1)/H$ be a coset, $H$ must be normal,
$H\cdot g=g\cdot H$, under the action defined in Eq.~(\ref{ac}).
This means that, for all $g\in ISO(2,1)$ and $h\in H$, there must exist an
$\bar h\in H$ such that $h\,g\,h^{-1}=\bar h^{-1}\,g\,\bar h$,
and we thus find that the only possible choices are
subgroups of the translation group $\real^3$, that is
$h_{_{L/R}}=(\id,y_{_{L/R}}^{\bar n})$, where $\bar n$ runs in a subset
of $\{0,1,2\}$ and $\id$ is the identity in $SO(2,1)$.
In this case, by inspecting the action (\ref{S}) one argues that
$\omega_\pm\equiv\xi_+ \equiv 0$, and $\xi_-^i\equiv 0$
iff the translation in the $i$ direction is not included in $H$.
The gauged action finally reads \cite{ch}
\be
S_g=-{\kappa\over 4}\,\int d^2\sigma\,\epsilon^{ijk}\,
\left(\partial_+\Lambda\,\Lambda^{-1}\right)_{ij}\,
\left(\partial_-v+\xi_-\right)_k
\ .
\label{S_g}
\ee
\par
For the ungauged action in Eq.~(\ref{S}) the equations of motion
$\delta_v S=0$, which follow from the variation $v\to v+\delta v$ with
$\delta v$ an infinitesimal 2+1 vector, lead to the conservation of the
six momentum currents on the light cone of the string world-sheet,
\be
\partial_- P_+^i=\partial_+ P_-^i=0
\ ,
\ee
where $P_+^i$ is given by $\bar P^i$ in Eq.~(\ref{P+}) with
$\sigma\to\sigma^+$ and $P_-^i$ by $P^i$ in Eq.~(\ref{P-})
with $\sigma\to\sigma^-$.
In the gauged case this variation must be supplemented by the condition
that the gauge field varies under an infinitesimal $H$
transformation according to
\be
\xi_-^{\bar n}\to \xi_-^{\bar n}-\partial_-(\delta v^{\bar n})
\ ,
\label{cond}
\ee
and from $\delta_v S_g=0$ one obtains
\be
\partial_-P_+^{i\not=\bar n}=0
\ ,
\label{dP}
\ee
so that only the currents $P_+^{i\not=\bar n}$ are still conserved.
\par
Similarly, from the variation $\Lambda\to\Lambda+\delta\Lambda$,
$\delta\Lambda=\Lambda\,\epsilon$ and $\delta v=\epsilon\,v$
with $\epsilon_{ij}=-\epsilon_{ji}$ an infinitesimal skewsymmetric
matrix, the equations $\delta_\epsilon S=0$ lead to the conservation
of the six angular momentum currents $J_-^i=J^i$ in Eq.~(\ref{J-})
with $\sigma\to\sigma^-$ and $J_+^i=\bar J^i$ in Eq.~(\ref{J+})
with $\sigma\to\sigma^+$.
When interpreted as components of the string angular momentum
in the target space-time, these currents are shown to include
a contribution of intrinsic (non orbital) spin \cite{stern}.
In the gauged case, by making use of Eqs.~(\ref{cond}) and (\ref{dP}),
one obtains
\be
\partial_+J_-^i=-\kappa\,\partial_+(\Lambda\,\xi_-)^i
\ ,
\label{dJ}
\ee
so that the currents $J_-^i$ couple to the gauge field.
\par
Since the gauge field is not dynamical, we are now free to choose
${\rm dim}\,H$ gauge conditions to be satisfied by the elements of
$ISO(2,1)/H$.
A natural choice is
\be
\xi^{\bar n}_-=-\partial_- v^{\bar n}
\label{gauge}
\ ,
\ee
so that the previous equations of motion become the same as
$\delta_v S_{eff}^{(\bar n)}=\delta_\epsilon S_{eff}^{(\bar n)}=0$
obtained by varying the effective action
\be
S_{eff}^{(\bar n)}=\int d^2\sigma\,\sum\limits_{k\not=\bar n} P_+^k\,
\partial_-v_k
\ ,
\label{S_eff}
\ee
where the sum runs over only the indices corresponding to the
translations not included in $H$.
\par
An explicit form for the effective action (\ref{S_eff}) can be obtained
by writing an $SO(2,1)$ matrix as a product of two rotations
(of angles $\alpha$ and $\gamma$) and a boost ($\beta$) \cite{ch},
\be
\Lambda^i_{\ j}&=&
\left[\begin{array}{ccc}
1 & 0 & 0 \\
0 & \cos\alpha & -\sin\alpha \\
0 & \sin\alpha & \cos\alpha
\end{array}\right]\,
\left[\begin{array}{ccc}
\cosh\beta & 0 & \sinh\beta \\
0 & 1 & 0 \\
\sinh\beta & 0 & \cosh\beta
\end{array}\right]\,
\left[\begin{array}{ccc}
1 & 0 & 0 \\
0 & \cos\gamma & -\sin\gamma \\
0 & \sin\gamma & \cos\gamma
\end{array}\right]
\ ,
\ee
which yields
\be
\begin{array}{l}
P^0_+=\strut\displaystyle{\kappa\over 2}\,
\left(\partial_+\alpha+\cosh\beta\,\partial_+\gamma\right)
\\
\\
P^1_+=\strut\displaystyle{\kappa\over 2}\,
\left(\cos\alpha\,\partial_+ \beta
+\sin\alpha\,\sinh\beta\,\partial_+\gamma\right)
\\
\\
P^2_+=\strut\displaystyle{\kappa\over 2}\,\left(
\sin\alpha\,\partial_+\beta
-\cos\alpha\,\sinh\beta\,\partial_+\gamma\right)
\ .
\end{array}
\label{P}
\ee
In the following we shall gauge a specific one-dimensional subgroup
which allows the number of degrees of freedom to be reduced from six to
four.
\section{Gauging the Time Translations}
\label{time}
We gauge the subgroup $H=\{(\id,y^0)\}$ of the translations in the time
direction.
This choice is peculiar, since no derivative of $\alpha$ occurs in
$P_+^1$ and $P_+^2$, and we can then rotate the variables $v^1$ and $v^2$
by an angle $-\alpha$ \cite{ch},
\be
\left[\begin{array}{c}
\partial_-\tilde v^1 \\
\partial_-\tilde v^2
\end{array}\right]
\equiv\left[\begin{array}{cc}
\cos\alpha & \sin\alpha \\
-\sin\alpha & \cos\alpha
\end{array}\right]\,
\left[\begin{array}{c}
\partial_-v^1 \\
\partial_-v^2
\end{array}\right]
\ .
\ee
This can be considered as an internal symmetry of the effective
theory which is used to further simplify the action in Eq.~(\ref{S_eff})
with $\bar n=0$ to the form \cite{alpha}
\be
S_{eff}^{(0)}={\kappa\over 2}\,\int d^2\sigma\,\left[
\partial_+\beta\,\partial_-\tilde v^1
-\sinh\beta\,\partial_+\gamma\,\partial_-\tilde v^2\right]
\ .
\label{S_4}
\ee
\par
In the following we shall find it more convenient to regard $\beta$,
$\gamma$, $\tilde v^2$ and $\tilde v^1$ as canonical (field) variables
by foliating the closed string world-sheet with circles of constant
time $\tau$ \cite{kuchar}.
Their conjugate momenta are then given by
\be
\begin{array}{l}
P^1\equiv\strut\displaystyle
{\delta S_{eff}^{(0)}\over\delta \partial_\tau{\beta}}=
\strut\displaystyle{\kappa\over 2}\,\partial_-\tilde v^1
\\
\\
P^2\equiv\strut\displaystyle
{\delta S_{eff}^{(0)}\over\delta \partial_\tau{\gamma}}=
-\strut\displaystyle{\kappa\over 2}\,\sinh\beta\,\partial_-\tilde v^2
\\
\\
P^3\equiv\strut\displaystyle
{\delta S_{eff}^{(0)}\over\delta \partial_\tau{\tilde v_2}}=
-\strut\displaystyle{\kappa\over 2}\,\sinh\beta\,\partial_+\gamma
=P_+^2(\alpha=0)
\\
\\
P^4\equiv\strut\displaystyle
{\delta S_{eff}^{(0)}\over\delta \partial_\tau{\tilde v_1}}=
\strut\displaystyle{\kappa\over 2}\,\partial_+\beta
=P_+^1(\alpha=0)
\ .
\end{array}
\ee
The above relations can be inverted to express the velocities in
terms of the momenta.
Therefore, the action (\ref{S_4}) does not contain any constraint and
its canonical structure can be analyzed straightforwardly.
The absence of constraints also signals the fact that all (explicit)
symmetries of the original model have been ``gauge fixed'' and
$\tilde v^1$, $\tilde v^2$, $\beta$ and $\gamma$ are physical degrees
of freedom.
We are then allowed to consider them all as target space-time coordinates
for the compactified string.
\par
From the point of view of the target space-time, with dimensionless
coordinates $X^1=\beta$, $X^2=\gamma$, $X^3=\tilde v^2$ and
$X^4=\tilde v^1$, the action (\ref{S_4}) can be written as \cite{ch2}
\be
S_{eff}^{(0)}&=&-{\kappa\over 2}\,\int d^2\sigma\,\left(\eta^{ab}+
\epsilon^{ab}\right)\,\left[\partial_a X^1\,\partial_b X^4
-F(X^1)\,\partial_a X^2\,\partial_b X^3\right]
\nonumber \\
&=&-{1\over 2}\,\int d^2\sigma\,\left[
\eta^{ab}\,G^{(4)}_{\mu\nu}\,
\partial_a X^\mu\,\partial_b X^\nu
+\epsilon^{ab}\,B^{(4)}_{\mu\nu}\,
\partial_a X^\mu\,\partial_b X^\nu\right]
\ ,
\label{S0eff}
\ee
where $F\equiv\sinh X^1$, $\eta^{ab}$ and $\epsilon^{ab}$ are, respectively,
the Minkowski tensor and the Levi-Civita symbol in two dimensions and
$\mu,\nu,\ldots=1,\ldots,4$.
By interpreting $\kappa=\ell_s^2$ as the square of the fundamental
(string) length,
the symmetric tensor ${\bol G}^{(4)}$ in the chosen reference frame
has components
\be
G^{(4)}_{\mu\nu}=\ell_s^2\,\left[\begin{array}{cccc}
0 & 0 & 0 & -1\\
0 & 0 & F & 0 \\
0 & F & 0 & 0 \\
-1 & 0 & 0 & 0
\end{array}\right]
\ee
and is the space-time metric with signature $2+2$ and
$G^{(4)}\equiv\det {\bol G}^{(4)}=\ell_s^8\,F^2$.
The antisymmetric tensor ${\bol B}^{(4)}$ has components
\be
B^{(4)}_{\mu\nu}=\ell_s^2\,\left[\begin{array}{cccc}
0 & 0 & 0 & -1 \\
0 & 0 & F & 0 \\
0 & -F & 0 & 0 \\
1 & 0 & 0 & 0
\end{array}\right]
\ee
and is the axion potential.
\par
The Euler-Lagrange equations of motion can be written as
\be
\begin{array}{l}
\delta_\beta S_{eff}^{(0)}=-\strut\displaystyle{\kappa\over 2}\,
\left[\partial_+\partial_- X^4
+\sqrt{1-F^2}\,\partial_+ X^2\,\partial_- X^3\right]=0
\\
\\
\delta_\gamma S_{eff}^{(0)}=-\strut\displaystyle{\kappa\over 2}\,
\partial_+\left(F\,\partial_- X^3\right)
=-\partial_+P^2=0
\\
\\
\delta_{\tilde v^2} S_{eff}^{(0)}=\strut\displaystyle{\kappa\over 2}\,
\partial_-\left(F\,\partial_+ X^2\right)
=-\partial_-P^3=0
\\
\\
\delta_{\tilde v^1} S_{eff}^{(0)}=-\strut\displaystyle{\kappa\over 2}\,
\partial_-\partial_+X^1=-\partial_-P^4=0
\ ,
\end{array}
\label{e4}
\ee
from which one sees that three of the canonical momenta ($P^2$, $P^3$ and
$P^4$) are conserved along (one of the two) null directions of the
world-sheet.
We also note that $X^1$ is a ``flat'' target space direction, since the
fourth of Eqs.~(\ref{e4}) is the free wave equation whose general solution
is given by
\be
X^1=X^1_L+X^1_R
\ ,
\ee
where the arbitrary functions $X^\mu_L=X^\mu_L(\sigma^+)$ stand for
left-moving and $X^\mu_R=X^\mu_R(\sigma^-)$ for right-moving waves.
\par
The system of Eqs.~(\ref{e4}) considerably simplifies for
zero canonical momentum modes along $X^2$ ($X^3=X^3_L$) or $X^3$
($X^2=X^2_R$), in which cases $X^4=X^4_L+X^4_R$.
When both $P^2$ and $P^3$ vanish one then has the simple solution
\be
\left\{\begin{array}{l}
X^1=X^1_L+X^1_R \\
\\
X^2=X^2_R \\
\\
X^3=X^3_L \\
\\
X^4=X^4_L+X^4_R
\ ,
\end{array}\right.
\label{sol}
\ee
which describes free wave modes in all of the four space-time directions.
More general solutions would instead describe wave modes which propagate
along a direction but couple with modes propagating in (some of) the
other directions.
\section{The final compactification}
\label{final}
Upon using the fact that $X^1=\beta$ is a flat direction for the
propagating string, we impose a further compactification condition
in order to eliminate $X^4=\tilde v^1$,
\be
e^{2\,\lambda}\,\partial_\pm x^1=\partial_\pm X^4
\ ,
\label{comp}
\ee
where $\lambda$ is, at present, an arbitrary function of $x^1\equiv X^1$.
We also define the two coordinates $x^0$ and $x^2$ according to
\be
\begin{array}{l}
\partial_+X^2=e^{\strut\displaystyle\rho}\,
\left(c_1\,\partial_+x^0+c_2^{-1}\,\partial_+x^2\right)
\\
\\
\partial_+X^3=e^{\strut\displaystyle\rho}\,
\left(c_1^{-1}\,\partial_-x^0-c_2\,\partial_-x^2\right)
\ ,
\end{array}
\label{lin1}
\ee
with $\rho=\rho(x^1)$ and $c_1$ and $c_2$ are non-zero real constants.
This reduces the action (\ref{S0eff}) to
\be
S_3&=&{\kappa\over 2}\,\int d^2\sigma\,\left[
e^{2\,\lambda}\,\partial_+ x^1\,\partial_- x^1
-e^{2\,\rho}\,F\,\left(\partial_+ x^0\,\partial_- x^0
-\partial_+ x^2\,\partial_- x^2\right)
+c_1\,c_2\,e^{2\,\rho}\,F\,
\left(\partial_+ x^0\,\partial_- x^2
-\partial_+ x^2\,\partial_- x^0\right)
\right]
\nonumber \\
&=&-{1\over 2}\,\int d^2\sigma\,\left[
\eta^{ab}\,G^{(3)}_{ij}\,
\partial_a x^i\,\partial_b x^j
+\epsilon^{ab}\,B^{(3)}_{ij}\,
\partial_a x^i\,\partial_b x^j\right]
\ ,
\label{s3}
\ee
where now the three-metric ${\bol G}^{(3)}$ has components
\be
G^{(3)}_{ij}=\ell_s^2\,{\rm diag}\,
\left[-e^{2\,\rho}\,F\,,\,e^{2\,\lambda}\,,\,e^{2\,\rho}\,F\right]
\ ,
\label{G3}
\ee
and signature $2+1$.
Further, the only non-vanishing independent component of the axion
potential ${\bol B}^{(3)}$,
\be
B^{(3)}_{02}=\ell_s^2\,c_1\,c_2\,e^{2\,\rho}\,F
\ ,
\label{B3}
\ee
does not depend on $\lambda$.
\par
We then observe that the axion field strength in three dimensions
must be proportional to the Levi-Civita (pseudo)tensor,
\be
H_{ijk}=\sqrt{-G^{(3)}}\,\epsilon_{ijk}\,{\cal H}
\ ,
\ee
where ${\cal H}={\cal H}(x^i)$ is a function of the space-time
coordinates to be determined from the field equations and
$\sqrt{-G^{(3)}}\equiv\sqrt{-\det{\bol G}^{(3)}}$ is the volume
element.
In the present case we have
\be
H_{012}=\partial_1 B^{(3)}_{20}=-\ell_s^2\,c_1\,c_2\,e^{2\,\rho}\,\left(
\sqrt{1-F^2}+2\,F\,\rho'\right)
\ ,
\ee
which yields
\be
{\cal H}=-{c_1\,c_2\over\ell_s}\,e^{-\lambda}\,
\left({\sqrt{1-F^2}\over F}+2\,\rho'\right)
\ .
\ee
\par
With respect to the particular solution (\ref{sol}), we note that the
condition (\ref{comp}) can be safely imposed only if $X^4_L=0$,
and then the compactification condition becomes
\be
e^{2\,\lambda(x_R^1)}\,\partial_- x_R^1=\partial_-X^4_R
\ ,
\label{comp1}
\ee
or $X^4_R=0$ and
\be
e^{2\,\lambda(x_L^1)}\,\partial_+ x_L^1=\partial_+X^4_L
\ .
\label{comp2}
\ee
\par
The functions $\rho$, $\lambda$ and the constants $c_1$, $c_2$ can then
be determined by noting that the low energy string action in three
dimensions is given by (we set $\ell_s=1$ henceforth) \cite{gsw}
\be
S_{low}=\int d^3x\,\sqrt{-G^{(3)}}\,e^{-2\,\phi}\,
\left[R+{4\over k}+4\,\nabla_k\phi\,\nabla^k\phi
-{1\over 12}\,H_{ijk}\,H^{ijk}\right]
\ ,
\ee
where $4/k$ is a cosmological constant, $R$ the scalar curvature,
$\nabla$ the covariant derivative with respect to the metric
${\bol G}^{(3)}$ and $\phi$ the dilaton.
On varying $S_{low}$ one obtains the field equations
\be
&&R_{ij}+2\,\nabla_i\nabla_j\phi
-{1\over 4}\,H_{ikl}\,H_j^{\ \ kl}=0
\label{f1}\\
&&\nabla_k\,\left(e^{-2\,\phi}\,H_{\ ij}^k\right)=0
\label{f2}\\
&&4\,\nabla_k\nabla^k\phi-4\,\nabla_k\phi\,\nabla^k\phi
+{4\over k}+R-{1\over 12}\,H_{ijk}\,H^{ijk}
=0
\ ,
\label{f3}
\ee
which must be satisfied by the metric (\ref{G3}) and the axion obtained
from the potential (\ref{B3}).
\subsection{Linear dilaton vacuum}
First we observe that for
\be
e^{-2\,\rho}=\pm F
\ ,
\ee
the metric (\ref{G3}) becomes the flat Minkowski metric
\be
ds^2=\mp\left(dx^0\right)^2+\left(dz_\pm\right)^2\pm\left(dx^2\right)^2
\ ,
\ee
where the upper signs correspond to $x^1>0$ ($F>0$) and the lower signs
to $x^1<0$ ($F<0$) and the new coordinate $z$ is determined by
\be
dz_\pm=e^{\lambda}\,dx^1={dx^1\over\sqrt{\pm F}}
\ .
\ee
Correspondingly ${\bol B}^{(3)}$ is constant and the axion vanishes, thus
the field equation (\ref{f2}) is trivially satisfied.
\par
The remaining Eqs.~(\ref{f1}) and (\ref{f3}) yield the following
expression for the dilaton field
\be
\phi=a+{x^0\over b}+{z_\pm\over c}+{x^2\over d}
\ ,
\ee
where the constant $a$ is arbitrary and the constants $b$, $c$ and $d$
must satisfy
\be
\pm{1\over b^2}-{1\over c^2}\mp{1\over d^2}=-{1\over k}
\ .
\ee
This solution represents a {\em linear dilaton vacuum}.
When $k\to\infty$ one of course obtains the trivial form for such a
vacuum with $\phi=a$.
\par
We finally observe that along $x^1=0$ there occurs a signature flip,
so that the roles of $x^0$ and $x^2$ as, respectively, a time coordinate
and a spatial coordinate are exchanged.
We shall find the same feature again in the following.
\subsection{Recovering AdS$_3$ and BTZ}
If we assume
\be
{\cal H}=-{2\over\ell}
\ ,
\ee
where $\ell$ is a constant, then the field equations (\ref{f1})-(\ref{f3})
are satisfied  by choosing $\rho=0$, $c_1\,c_2=1$ and
\be
e^{2\,\lambda}={\ell^2\over 4}\,\coth^2 x^1
\ ,
\label{const}
\ee
which yields
\be
X^4={\ell\over 2}\,\left(x^1-\coth x^1\right)+X_0^4
\ ,
\ee
with $X_0^4$ an integration constant.
It then follows that the compactification we are employing is indeed
singular, since $X^4\sim \mp1/x^1$ for $x^1\to 0^\pm$, which means that
we are mapping vanishing boosts along $v^1$ (parameterized by $\beta$)
into infinite translations along $\tilde v^1$.
For this reason we tentatively consider the range of $x^1=\beta$ as
divided into the two (disjoint) half lines $x^1>0$ and $x^1<0$.
\par
It is indeed possible to show that this partition of the range of $\beta$
has a natural interpretation in terms of the space-time manifold.
In fact, the choice (\ref{const}) reduces Eq.~(\ref{s3}) to the action for
a string propagating in the three-dimensional AdS space-time.
This can be seen, {\em e.g.}, by defining new (dimensionless) coordinates
$r_\pm\in\real$ such that
\be
\begin{array}{lrl}
r_+=\ln(+\sinh x^1)
&\ \ \ \ {\rm for}& x^1>0
\\
& & \\
r_-=\ln(-\sinh x^1)
&\ \ \ \ {\rm for}& x^1<0
\ .
\end{array}
\ee
The metric
\be
ds^2=\sinh x^1\,\left[
\left(dx^2\right)^2-\left(dx^0\right)^2\right]
+{\ell^2\over 4}\,\coth^2 x^1\,\left(dx^1\right)^2
\ee
then becomes
\be
ds^2=\pm\,e^{\strut\displaystyle{r_\pm}}\,\left[
\left(dx^2\right)^2-\left(dx^0\right)^2\right]
+{\ell^2\over 4}\,\left(dr_\pm\right)^2
\ ,
\label{2ads}
\ee
where the equality holds with the plus sign for $x^1>0$ and with the minus
sign for $x^1<0$.
The expression in Eq.~(\ref{2ads}) is one of the standard forms for AdS$_3$
with $x^0$ (or $x^2$) playing the role of time and $r_+$ and $x^2$
(or $r_-$ and $x^0$) of spatial coordinates.
This can perhaps be more easily recognized if one defines a coordinate
\be
z=\exp\left(-{r_\pm\over2}\right)
\ ,
\ee
and obtains
\be
ds^2={\pm\left[\left(dx^2\right)^2-\left(dx^0\right)^2\right]
+\ell^2\,\left(dz\right)^2\over z^2}
\ ,
\label{ads}
\ee
which describes the half of (one of the two) AdS$_3$ with $z>0$
(the half $z<0$ being given by a definition of $z$ with the opposite
sign).
The BTZ black hole is then obtained by the usual periodicity condition
imposed on the coordinates \cite{BTZ,brill}.
\par
It then follows that the metric (\ref{2ads}) we have found simultaneously
describes two copies of (half of) AdS$_3$ and $x^1=0$ again plays the role
of a boundary at $r_+=r_-=-\infty$ across which the signature of the metric
flips.
In fact this is the standard AdS horizon at $|z|=+\infty$, while the
time-like infinity is at $z=0$, and the scalar curvature,
\be
R=-{6\over\ell^2}
\ ,
\ee
is a negative (regular) constant everywhere.
Of course, the metrics (\ref{2ads}) and (\ref{ads}) solve the field
equations (\ref{f1})-(\ref{f3}) provided $\phi$ is also a constant and
$k=\ell^2$ \cite{horowitz}.
\par
We conclude this part by noting that the solution (\ref{sol})
places further restrictions on the propagating modes, since $F$ can then
be a function of either $x^1_L$ or $x^1_R$, but not of both
[see Eqs.~(\ref{comp1} and (\ref{comp2})].
This selects out a subclass of solutions with only left- (or right-)
movers along $x^1$ and both kinds of waves along $x^0$ and $x^3$.
\section{Other solutions}
\label{other}
Various other solutions to the field equations (\ref{f1})-(\ref{f3})
can be found for a non-constant dilaton.
\subsection{First example}
\label{first}
Let us consider the metric
\be
G^{(3)}_{ij}={\rm diag}\,
\left[-F\,,\,e^{2\,\lambda}\,,\,F\right]
\ ,
\label{G31}
\ee
where now $\lambda=\lambda(x^2)$ and similarly for the dilaton
$\phi=\phi(x^2)$.
This metric can be obtained from the action (\ref{S0eff}) by applying
again a nonlinear transformation of the form (\ref{comp})-(\ref{lin1}),
\be
\partial_-X^4&=&e^{2\,\lambda}\,\partial_-x^1
\nonumber \\
\partial_+X^2 &=&
e^{\rho}\,\left[c_1\,\partial_+x^0+c_2^{-1}\,\partial_+x^2\right]
\nonumber\\
\partial_-X^3&=&
e^{\rho}\,\left[c_1^{-1}\,\partial_-x^0-c_2\,\partial_-x^2\right]
\ ,
\ee
where $c_1$ and $c_2$ are constants and $\rho=\rho(x^1)$.
If we choose $\rho$ such that
\be
e^{-2\,\rho}=F
\ ,
\ee
the $\sinh(x^1)$ term in Eq.~(\ref{S_4}) is cancelled, and the axion
potential ${\bol B}^{(3)}$ is constant in our model, so the axion field
strength ${\cal H}$ is zero.
Substituting this form for ${\bol G}^{(3)}$ into the field equations
(with $k=1$) allows us to determine $\lambda$ and leads to the invariant
line element
\be
ds^2=-(dx^0)^2+\coth^2(x^2)\,(dx^1)^2+(dx^2)^2
\ .
\label{S1}
\ee
The dilaton in this case is
\be
\phi=C_\phi-\ln\left(\sinh(x^2)\right)
\ ,
\ee
where $C_\phi$ is an integration constant and the domain of validity of the
solution is $x^2>0$ (which we call region II).
This metric is ``asymptotically flat'', in the sense that it converges to
the Minkowski metric for $x^2\to\infty$.
The Ricci scalar is negative and diverges at the origin ({\em i.e.}, for
$x^2=0$),
\be
R=-{4\over\sinh^2(x^2)}
\ .
\label{R1}
\ee
\par
The change of variables \cite{theta}
\be
t=x^0\, ,\
r=\coth(x^2)\, ,\
\theta=x^1
\ ,
\ee
brings the invariant line element into the form
\be
ds^2=-dt^2+{dr^2\over{(r^2-1)^2}}+r^2\,d\theta^2
\ ,
\label{g1_2}
\ee
which shows explicitly the cylindrical symmetry.
The radial coordinate is understood to be $r>1$, according to the above
definition of region II, and the dilaton is written as
\be
\phi_{II}=C_\phi+{1\over 2}\,\ln(r^2-1)
\ .
\ee
In the form (\ref{g1_2}), the metric can also be extended to the region
$0\le r<1$ (region I), where the Ricci scalar,
\be
R=4\,\left(1-r^2\right)
\ ,
\ee
is positive and regular everywhere and the dilaton becomes
\be
\phi_I=C_\phi+{1\over 2}\,\ln(1-r^2)
\ .
\ee
\par
A plot of an angular sector of the ``lifted surface'' \cite{mtw2},
adapted so as to include the origin of region I, is shown in Fig.~1.
With this choice a new singularity appears at $r=1$, where the surface
has diverging slope and would extend to unlimited height (the plot is of
course truncated along the vertical axis).
This simply represents the fact that the asymptotically flat region
($r\to 1$) is at an infinite proper distance ($\sim\ln|r-1|$)
both from the origin $r=0$ of region I and from the singularity
$r\to\infty$ ($x^2=0$) of region II.
In fact, the Ricci scalar actually vanishes along $r=1$ and the only real
singularity is at $r\to\infty$.
\par
The real singularity in region II is not accessible from region I.
In particular, by solving the equation of radial null geodesics,
\be
{d^2r\over d\tau^2}
-{2\,r\over r^2-1}\,\left({dr\over d\tau}\right)^2=0
\ ,
\ee
where $\tau$ is an affine parameter, one finds (near $r=1$ and with
$\tau\gg 0$)
\be
r\sim 1\pm e^{-\tau}
\ ,
\ee
where the minus (plus) sign is for geodesics starting in
region I (II).
Such trajectories define the light cones in regions I and II and
therefore show that the two regions are causally disconnected.
\begin{figure}
\centerline{\epsfysize=200pt\epsfbox{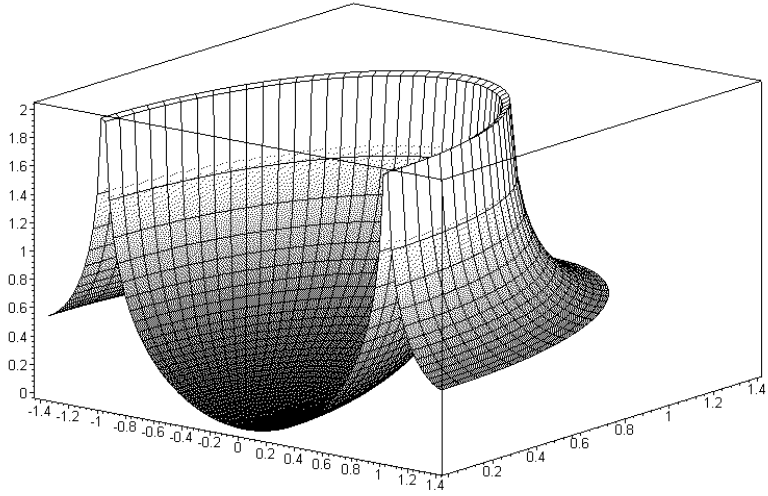}}
\caption{The sector $0\le\theta\le\pi$ of the lifted surface for the metric
(\ref{g1_2}) extended to all positive values of $r$.}
\end{figure}
\subsection{Second example}
\label{second}
Let us now consider the metric
\be
G^{(3)}_{ij}={\rm diag}\,
\left[-e^{-2\,\rho},\,e^{2\,\rho}\,,\,e^{2\,\rho}\right]
\ ,
\label{G32}
\ee
where now $\rho=\rho(x^1)$ and $\phi=\phi(x^1)$.
This metric results from the nonlinear coordinate transformation
\be
\partial_-X^4&=&e^{2\,\rho}\,\partial_-x^1
\nonumber \\
\partial_+X^2&=&
F^{-1/2}\,\left[c_1\,e^{-\rho}\,\partial_+x^0
+c_2^{-1}\,e^{\rho}\,\partial_+x^2\right]
\nonumber\\
\partial_-X^3&=&
F^{-1/2}\,\left[c_1^{-1}\,e^{-\rho}\,\partial_-x^0
-c_2\,e^{\rho}\,\partial_-x^2\right]
\ .
\ee
The transformation of coordinates used to obtain this form for
${\bol G^{(3)}}$ in our model insures that the axion potential is again
constant.
Eliminating $x^1$ in favor of $\rho$ and solving the field equations again
with $k=1$, we find for the invariant line element
\be
ds^2=-e^{-2\,\rho}\,(dx^0)^2+{4\,d\rho^2\over\sinh^2(\sqrt{2}\,\rho)}
+e^{2\,\rho}\,(dx^2)^2
\ ,
\ee
and the dilaton is
\be
\phi=C_\phi-{1\over 2}\,\ln\left(\sinh(\sqrt{2}\,\rho)\right)
\ ,
\ee
with $C_\phi$ the usual integration constant.
\par
The change of variables \cite{theta}
\be
t=x^0\, ,\
r=e^\rho\, ,\
\theta=x^2
\ ,
\ee
which is well defined for $r>0$, gives the invariant line element the
manifestly cylindrically symmetric form
\be
ds^2=-{dt^2\over r^2}+{dr^2\over r^2\,(r^{\sqrt{2}}-r^{-\sqrt{2}})^2}
+r^2\,d\theta^2
\ ,
\label{g2_2}
\ee
and the dilaton can now be written as
\be
\phi=C_\phi-{1\over 2}\,\ln\left(r^{\sqrt{2}}-r^{-\sqrt{2}}\right)
\ .
\ee
The Ricci scalar is everywhere negative,
\be
R=-\left(r^{\sqrt{2}}-r^{-\sqrt{2}}\right)^2
\ ,
\ee
has essential singularities at both $r=0$ and $r\to\infty$ and
vanishes along the circle $r=1$.
This implies a similarity with the previous metric (\ref{g1_2}), namely
one can define a region I for $0<r<1$ and a region II for $r>1$.
The main difference is then that region I also contains a real
singularity at $r=0$.
\subsection{T-dual solutions}
We conclude this Section by noting the in both solutions above,
there are two isometric coordinates, to wit $t$ and $\theta$.
Therefore, one can generate new solutions by employing T-duality
\cite{buscher,ch2}.
In particular, we shall T-dualize with respect to one coordinate
at a time, and denote the fields of the dual solutions with a tilde.
We also denote by $B$ the only non-vanishing component of the axion
potential, $B^{(3)}_{02}$, which is constant in all solutions
considered.
\par
For the solution in Section~\ref{first}, the non-vanishing component of the
axion potential in the coordinate system $(t,r,\theta)$ is given by
\be
B^{(3)}_{tr}={B\over 1-r^2}
\ .
\ee
On dualizing the metric (\ref{g1_2}) with respect to $t$ then yields the
non-diagonal line element
\be
\tilde{ds}^2=-dt^2+{1-B^2\over(r^2-1)^2}\,dr^2+{B\,dt\,dr\over r^2-1}
+r^2\,d\theta^2
\ ,
\ee
which solves the field equations with a vanishing axion potential,
$\tilde{\bol B}^{(3)}=0$, and an unchanged dilaton field, $\tilde\phi=\phi$
\cite{buscher}.
\par
Dualizing with respect to $\theta$ instead leaves the metric unaffected,
as can be seen by switching to the new radial coordinate $R=r^{-1}$ after
applying the dual relations \cite{buscher}, but yields
$\tilde{\bol B}^{(3)}=0$ and a shifted dilaton field,
$\tilde\phi=\phi+\ln(R)$.
\par
The duals of the metric (\ref{g2_2}) and of the axion potential
$B^{(3)}_{t\theta}=B$ of Section~\ref{second} with respect to $t$
are given by
\be
\tilde{ds}^2=-r^2\,dt^2+{dr^2\over r^2\,(r^{\sqrt{2}}-r^{-\sqrt{2}})^2}
-B\,r^2\,dt\,d\theta+r^2\,(1-B^2)\,d\theta^2
\ee
and $\tilde{\bol B}^{(3)}=0$, with the dilaton $\tilde\phi=\phi+\ln(r)$.
The metric above represent a rotating space-time, since the off-diagonal
term $\tilde G^{(3)}_{t\theta}\not=0$ (for $B\not=0$).
\par
Dualizing with respect to $\theta$ and defining $R=r^{-1}$ gives
\be
\tilde{ds}^2=-R^2\,(1-B^2)^2\,dt^2
+{dR^2\over R^2\,(R^{\sqrt{2}}-R^{-\sqrt{2}})^2}
-B\,R^2\,dt\,d\theta+R^2\,d\theta^2
\ ,
\ee
$\tilde{\bol B}^{(3)}=0$ and $\tilde\phi=\phi+\ln(R)$.
Again this represents a rotating space-time.
\par
In three out of four cases above the presence of a non-vanishing axion
potential, although it corresponds to zero field strength, affects the
metric field in a non-trivial manner.
The axion potential is in fact always absorbed into the dual metric and
dilaton fields and sometimes generates off-diagonal terms and rotation.
\section{Conclusions}
\label{end}
Starting from the six parameter group $ISO(2,1)$ we have, by using various
types of compactification (gauge fixing, internal symmetries and coordinate
identification), reduced the original action, which describes a spinless
string moving on a curved six-dimensional background, to a string propagating
on either a flat (Minkowski) background with a linear dilaton or on AdS
space-time with a constant dilaton field.
If the fields satisfying the equations obtained from the low energy
effective string action are restricted to be functions of a single variable
(in our case one of the boost parameters from the original Poincar\'e group),
the fields are so tightly constrained that there are apparently only two
possible solutions with a trivial dilaton.
\par
The original goal of this work was to find a three-dimensional black hole
other than the BTZ black hole by starting from a model of string propagation
on a group manifold different from the $SL(2,\real)$ manifold.
This goal has not been realized, but the tactic has resulted in a relatively
simple form for the compactified Lagrangian, allowing us to recover the
space-time of AdS$_3$ (and BTZ) and to obtain solutions of the field
equations we might not otherwise have been able to attain.
\par
The fact that AdS$_3$ can be related to the (non-semisimple)
three-dimensional Poincar\'e group might be surprising at first sight.
However, one can consider the following general argument:
The natural group of symmetry of AdS$_3$, that is the semisimple group
$SL(2,\real)$, is contained within $SL(2,C)$ which, in turn, is
isomorphic to $SO(3,1)$.
Moreover, the Lie algebra of the group $ISO(2,1)$ can be reached from
the Lie algebra of the (semisimple) group $SO(3,1)$ by means of a
transformation called {\em contraction}
(see, {\em e.g.}, Ref.~\cite{gilmore}).
One can therefore conclude that the sequence of operations we have
performed reproduces the (local) effect of an {\em expansion}
(roughly, the opposite of a contraction \cite{gilmore}) from the coset
$ISO(2,1)/\real$ to $SL(2,\real)$.
\par
Other such formal constructions can be envisoned, and might turn out to be
useful in the search for new solutions, as we have shown in
Section~\ref{other}.
Whether or not the BTZ black hole is the unique one in three-dimensional
space-time remains an open question, and so, therefore, is the question of
whether or not another exact three-dimensional black hole solution to string
theory exists.
\acknowledgments
This work was supported in part by the U.S. Department of Energy
under Grant No. DE-FG02-96ER40967 and by the NATO grant No. CRG 973052.

\begin{references}
%
\bibitem{bh}
For a recent review see A. Celotti, S.C. Miller and D.W. Sciama,
Class. Quantum Grav. {\bf 16}, A3 (1999).
%
\bibitem{gsw}
M.B. Green, J.H. Schwarz and E. Witten, {\em Superstring theory},
Cambridge Univ. Press, Cambridge, England (1987).
%
\bibitem{duff}
M. Duff, R. Khuri and J. Lu, Phys. Rep. {\bf 259}, 213 (1995).
%
\bibitem{horowitz}
G.T. Horowitz and D.L. Welch, Phys. Rev. Lett. {\bf 71}, 328 (1993).
%
\bibitem{witten2}
E. Witten, Nucl. Phys. {\bf B311} (1988) 46; {\bf B323}, 113 (1989);
H.-J. Matschull, Class. Quantum Grav. {\bf 16}, 2599 (1999).
%
\bibitem{BTZ}
M. Ba\~nados, C. Teitelboim and J. Zanelli, Phys. Rev. Lett.
{\bf 69}, 1849 (1992);
M. Ba\~nados, M. Henneaux, C. Teitelboim and J. Zanelli,
Phys. Rev. D {\bf 48}, 1506 (1993).
%
\bibitem{brill}
D. Brill, {\em Black holes and wormholes in 2+1 dimensions}, to appear
in the {\em Proceedings of the 2$^{nd}$ Samos meeting on cosmology,
geometry and relativity}, preprint gr-qc/9904083.
%
\bibitem{steif}
A. Steif, Phys. Rev. D {\bf 49}, 585 (1994).
%
\bibitem{satoh}
Y. Satoh, Ph.D. thesis (unpublished), hep-th/9705209.
%
\bibitem{maldacena}
J. Maldacena, Adv. Theor. Math. Phys. {\bf 2}, 231 (1998);
E. Witten, Adv. Theor. Math. Phys. {\bf 2}, 253 (1998).
%
\bibitem{sen}
D. Birmingham and S. Sen, {\em Gott time machines, BTZ black hole
formation, and Choptuik scaling}, preprint hep-th/9908150.
%
\bibitem{hawking}
S.W. Hawking, Nature (London) 248, 30 (1974);
Comm. Math. Phys. 43, 199 (1975).
%
\bibitem{birrell}
N.D. Birrell and P.C.W. Davies, {\em Quantum fields in curved space},
Cambridge Univ. Press, Cambridge, England (1982).
%
\bibitem{proof}
An algebraic proof of the the correspondence for quantum field theories
in AdS has been recently claimed; see, {\em e.g.},
K.-H. Rehren, {\em A proof of the AdS-CFT correspondence},
hep-th/9910074.
%
\bibitem{mfd}
R. Casadio and B. Harms, Phys. Rev. D {\bf 58}, 044014 (1998);
R. Casadio, B. Harms and Y. Leblanc, {\em Semiclassical quantization
on black hole space-time}, in {\em Proceedings of the 8$^{th}$
Marcel Grossman meeting on general relativity}, edited by T. Piran
and R. Ruffini, World Scientific, Singapore (1999).
%
\bibitem{stern}
P. Salomonson, B. Skagerstam and A. Stern, Nucl. Phys. {\bf B347},
769 (1990);
A. Stern in {\em Superstrings and Particle theory}, ed.
L. Clavelli and B. Harms, World Scientific, Singapore (1990).
%
\bibitem{ch}
R. Casadio and B. Harms, Phys. Lett. {\bf B 389}, 243 (1996).
%
\bibitem{ch2}
R. Casadio and B. Harms, Phys. Rev. D {\bf 57}, 7507 (1998).
%
\bibitem{mtw}
C.W. Misner, K.S Thorne and J.A. Wheeler,
{\it Gravitation}, W.H. Freeman and Co., San Francisco
(1973).
%
\bibitem{witten}
E. Witten, Comm. Math. Phys. {\bf 92}, 455 (1984).
%
\bibitem{chiral}
This gives half the canonical structure, corresponding to the right
moving string modes.
In fact, one can equivalently map $\sigma^-\to\tau$ and
$\sigma^+\to\sigma$ and obtain a duplicate of the Poisson algebra
(\ref{P,P})-(\ref{J,P}) for the (independent) left movers
\cite{witten,stern}.
%
\bibitem{bound}
The existence of ghosts is due to the boundary conditions that are
imposed on the spin connection of the topological theory at
$\partial{\cal M}$ which render the reparametrization invariance at
$\partial{\cal M}$ anomalous \cite{stern}.
%
\bibitem{stern2}
A. Stern, Nucl. Phys. {\bf B482}, 305 (1996).
%
\bibitem{alpha}
The same result can be obtained by setting $\alpha=0$ in Eq.~(\ref{S_eff})
with $\bar n=0$.
%
\bibitem{kuchar}
K.V. Kuchar and C.G. Torre, J. Math. Phys. {\bf 30}, 1769 (1989).
%
\bibitem{mtw2}
See, {\em e.g.}, Ref.~\cite{mtw}, pag.~614.
%
\bibitem{theta}
This change of variables assumes a compactification along the isometric
coordinate $\theta$ such that $\theta\sim\theta+2\,\pi$.
%
\bibitem{buscher}
T.H. Buscher, Phys. Lett. {\bf B 194}, 51 (1987); {\bf B 201}, 466 (1988).
%
\bibitem{gilmore}
R. Gilmore, {\em Lie groups, Lie algebras and some of their applications},
Chap.~10, J. Wiley and Sons, New York (1974).
%
\end{references}
\end{document}